\documentclass{PoS}
\usepackage{booktabs,graphicx,siunitx,xspace}
\usepackage{subfigure}
\usepackage{wrapfig}
\usepackage{epsfig}
\usepackage{epstopdf}

\newcommand*\aap{A\&A}

\newcommand*\aj{AJ}
\newcommand*\ao{Appl.~Opt.}

\newcommand*\apj{ApJ}
\newcommand*\apjl{ApJ}

\newcommand*\apjs{ApJS}

\newcommand*\araa{ARA\&A}

\newcommand*\mnras{MNRAS}

\newcommand*\nat{Nature}

\newcommand*\prd{Phys.~Rev.~D}

\let\oldbibliography\thebibliography
\renewcommand{\thebibliography}[1]{%
  \oldbibliography{#1}%
  \setlength{\itemsep}{0pt}%
}

\title{Positron Annihilation in the Milky Way and beyond}

\ShortTitle{Positron annihilation in the Milky Way and beyond}

\author{\speaker{Thomas Siegert}\\
        Max-Planck-Institute for extraterrestrial Physics, D-85748 Garching, Germany\\
        E-mail: \email{tsiegert@mpe.mpg.de}}

\abstract{The electron-positron annihilation gamma-ray signal at 511~keV in the Milky Way is investigated towards a possible dark matter interpretation. If all bulge positrons were created by dark matter particle annihilation, the satellite galaxies of the Milky Way, apparently being dominated by dark matter, should also show measurable 511~keV signals. Using INTEGRAL/SPI, we test for emission in 39 neighbouring dwarf satellite galaxies, and found a consistent trend against a dark matter scenario. One galaxy, Reticulum II, shows up as a strong source of annihilation emission, which we interpret as the presence of a microquasar, ejecting pair-plasma into the galaxy's interstellar medium.}

\FullConference{11th INTEGRAL Conference Gamma-Ray Astrophysics in Multi-Wavelength Perspective,\\
		10-14 October 2016\\
		Amsterdam, The Netherlands}

\begin{document}

\section{Context}
For more than 40 years, gamma-ray astronomers measure a strong and extended signal from the centre of the Milky Way which they cannot explain thoroughly: the electron-positron annihilation ($\mathrm{e^-,e^+}$) signal at photon energies around 511~keV~\cite{Haymes1969_511,Johnson1972_511,Leventhal1978_511}. Our current knowledge in this context is restricted to where the positrons annihilate (gamma-ray morphology~\cite{Purcell1997_511,Knoedlseder2005_511,Bouchet2010_positron,Skinner2014_511,Siegert2016_511}), and how they annihilate (gamma-ray spectral shape~\cite{Leventhal1978_511,Jean2006_511,Churazov2005_511,Churazov2011_511,Siegert2016_511}). However, we do not know where and how the $\mathrm{e^+}$s are created, i.e. which astrophysical sources are the major contributors, and whether there is only one type of sources or whether there are more. A review of possible scenarios and sources is given in Prantzos et al. (2011)~\cite{Prantzos2011_511}, while Siegert, T. (2017) provides a first complete comprehension of $\mathrm{e^+}$ sources, based on observational evidences~\cite{Siegert2017_PhD}. In principle, almost all astrophysical objects and events, be it stars or black holes, supernovae or cosmic rays, produce $\mathrm{e^+}$s or $\mathrm{e^+e^-}$-pairs. There is, in fact, a large wealth of possible quantum-processes which create $\mathrm{e^+}$s, e.g. by $\mathrm{\beta^+}$-decay of radioactive nuclei, pair-creation in electro-magnetic fields, or through photon-photon interactions. This means, there are too many possible explanations to naively pick one as \textit{the} answer.

The 511~keV morphology in the Galaxy is dominated by largely-extended diffuse emission from the direction of the galactic centre, called the bulge\footnote{This 511~keV does not necessarily have something to do with the infrared bulge, for example, but is only named this way because of the morphological similarities.}. This bulge can be characterised by at least two 2D-Gaussian-shaped components with radial extents of $\sim 6^{\circ}$ (FWHM, offset to $l \approx -1^{\circ}$), and $\sim 20^{\circ}$ (FWHM, centred on $l=0^{\circ}$). Skinner et al. (2014) found, that a third, point-like component around $(l/b) = (0^{\circ}/0^{\circ})$ is also adequate to describe the INTEGRAL/SPI data~\cite{Skinner2014_511}. Siegert et al. (2016a) tested this hypothesis and found a $5\sigma$-improvement over their base-line model without that source, and could also provide a first spectrum of this feature~\cite{Siegert2016_511}. In addition to the bulge, several authors found a low surface-brightness disk~\cite{Knoedlseder2005_511,Weidenspointner2008a_511,Bouchet2010_511,Skinner2014_511,Siegert2016_511}. Due to increased exposure off the galactic centre after ten years of INTEGRAL observations, the longitudinal and latitudinal extents could be determined to $\sigma_l = 60^{+10}_{-5}~\mathrm{deg}$ and $\sigma_b = 10.5^{+2.5}_{-1.5}~\mathrm{deg}$~\cite{Siegert2016_511}, favouring an old stellar population as dominant contributors.

The total 511~keV line flux, as measured from the high-resolution gamma-ray spectra obtained with SPI, is about $1.0 \times 10^{-3}~\mathrm{ph~cm^{-2}~s^{-1}}$ for the bulge, and $\sim 1.7\times 10^{-3}~\mathrm{ph~cm^{-2}~s^{-1}}$ for the disk~\cite{Siegert2016_511}. The annihilation emission is detected above a background-only model with more than $50\sigma$ significance in the bulge, and $12\sigma$ in the disk. In addition to the line fluxes, we are able to determine details of the spectral shapes, e.g. line widths ($2.6\pm0.2~\mathrm{keV}$) and centroids ($511.09\pm0.08~\mathrm{keV}$), and also the contribution from an annihilation continuum below 511~keV from the "decay" of positronium atoms, a short-lived bound state between $\mathrm{e^-}$s and $\mathrm{e^+}$s. These spectral parameters are similar to within $\approx 2 \sigma$ for all components needed to describe the Milky Way. By combining the fitted parameters and underlying a model description of the interstellar medium, e.g. assuming a single phase medium, pure hydrogen, dust free, and as a function of temperature and ionisation state~\cite{Churazov2005_511,Churazov2011_511}, we can derive the annihilation conditions. The preferred medium for $\mathrm{e^+}$s to annihilate in, is then moderately warm ($T=7000$-$40000~\mathrm{K}$), and partly ionised ($x=2$-$7\%$). The dominant annihilation channel - the "how" - is found to be charge exchange reactions with mostly neutral hydrogen.

The kinetic energy of the $\mathrm{e^+}$s we see to annihilate cannot be much larger than $\sim1$~keV, as otherwise the 511~keV line would be broader, and an additional annihilation continuum \textit{above} 511~keV, up to the kinetic energy of the $\mathrm{e^+}$-population, would be seen~\cite{Sizun2006_511,Beacom2006_511,Churazov2011_511}. If this type of model is applied to all the different emission components' spectra, actually a large variety of annihilation conditions is possible: from peculiar sampling of dust-dominated regions in the disk to annihilation in fully ionised gas in the centre of the Galaxy. However, almost all quantum-processes, which create $\mathrm{e^+}$s, eject them at relativistic energies. Thus, there are (series of) processes involved which must decelerate $\mathrm{e^+}$s along their ways from the astrophysical production sites to the annihilation sites. The propagation of $\mathrm{e^+}$s is an inevitable consequence, and the gamma-ray morphology that is observed with INTEGRAL/SPI is not necessarily the source morphology.

By means of theoretical calculations and assumptions regarding the regions in which the $\mathrm{e^+}$s are created, we can estimate how long they survive before annihilating with $\mathrm{e^-}$, predominantly from hydrogen. Various authors simulated diffusion-like transport processes of $\mathrm{e^+}$s in the interstellar medium~\cite{Guessoum1991_511ISM,Guessoum2005_511,Jean2009_511ISM,Alexis2014_511ISM}. Depending on the density of the interstellar gas, $\mathrm{e^+}$s might survive between $\sim0.1$ and $\sim10$~Myrs, corresponding to distances between $\sim 0.1$ and $\gtrsim 1.0$~kpc, before annihilation. The true interstellar gas conditions near the candidate sources are not known, so that the escape from those sources may be questionable in general. However, if the source is omnipresent, rather than punctual, the smooth diffuse and bulge-dominated emission morphology, together with the annihilation conditions (and also the disputable escape) could readily be explained by a bulge component with its own source: dark matter.

\section{A dark matter solution?}

The above scenario has already been pursued in times when there was only emission confidently seen from the galactic centre~\cite{Boehm2004_dm,Cordier2004_511dm,Ascasibar2006_511dm,Huh2008_511dm,Boehm2008_dm}. In fact, the 511~keV bulge components are very reminiscent of a dark matter halo density-profile. Skinner et al. (2014) found that a single dark matter halo profile,

\begin{equation}
\rho_{DM}(r) = \frac{\rho_0}{(r/R)^{\gamma}\left[1 + (r/R)^{\alpha}\right]^{(\beta-\gamma)/\alpha}}\mathrm{,}
\label{eq:dm_profile}
\end{equation}

with $\alpha=1$, $\beta=3$, $\gamma=1$ (Navarro-Frenk-White profile), and $R=20$~kpc can replace all three bulge components by statistical means~\cite{Skinner2014_511}. In general, there are three possible processes how dark matter particles (e.g. axions, WIMPs, neutralinos, ...) could produce $\mathrm{e^+e^-}$-pairs: decay~\cite{Hooper2004_dm511axion}, annihilation~\cite{Boehm2004_dm,Hooper2004_dm}, or de-excitation~\cite{Finkbeiner2007_dm511deex}. In the emission morphology, this is incorporated as an exponent $n$ to the density profile: $F_{511} \propto \rho_{DM}(r)^n$, where $n=1$ corresponds to a one-particle process, i.e. decay, and $n=2$ to a two-particle process, i.e. self-annihilation or de-excitation. Only $n=2$ is possible for the SPI 511~keV data~\cite{Skinner2014_511}. The possible shift of the 511~keV emission towards negative longitudes may also be explained, as simulations~\cite{Kuhlen2013_dm} showed that the "position of the central dark matter density peak may be expected to differ from the dynamical centre of the Galaxy by several hundred parsecs", i.e. about $1^{\circ}$ in longitude. 

\pagebreak

\subsection{How to test the dark matter scenario?}

\begin{figure}[!ht]%
\centering
\includegraphics[width=0.75\columnwidth]{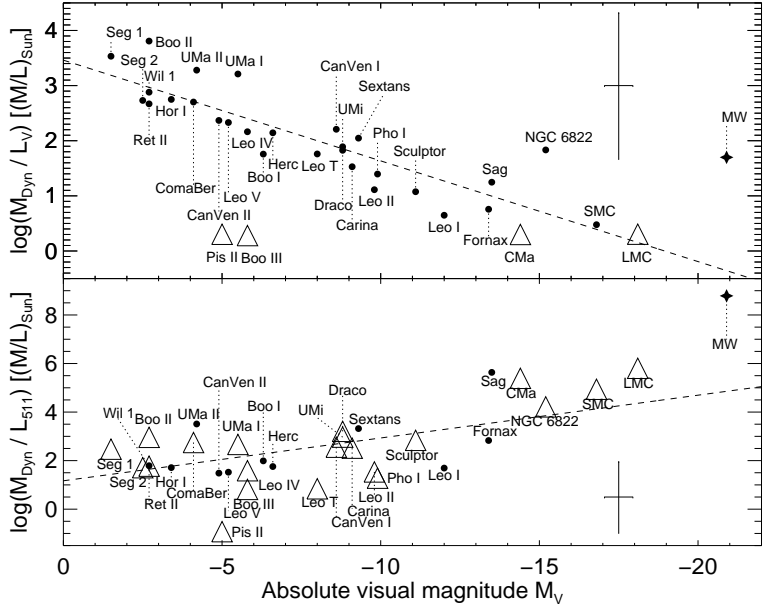}%
\caption{Mass-to-light ratio for dwarf satellite galaxies of the Milky Way. Top panel: Ratio between the dynamical mass and the V-band luminosity, $\Upsilon_V$, normalised to solar units, against the absolute V-band magnitude $M_V$. Bottom panel: Ratio between the dynamical mass and the 511~keV luminosity, $\Upsilon_{511}$, against the respective V-band magnitudes. For a dark matter origin of the 511~keV emission, a similar trend for $\Upsilon_{511}$ as for $\Upsilon_V$ would be expected, but which is not seen. See text for details. From Siegert et al. (2016b)~\cite{Siegert2016_dsph}.}%
\label{fig:masstolight}%
\end{figure}

If the entire bulge annihilation radiation originates from dark matter particles, either due to annihilation or de-excitation\footnote{For brevity, we use "annihilation" synonymously for annihilation and de-excitation, i.e. an $n=2$ process.}, the dwarf satellite galaxies of the Milky Way should also be seen in 511~keV~\cite{Hooper2004_dm,Simon2007_dm,Strigari2008_dm}, because they are apparently dominated by dark matter~\cite{Mateo1998_dsphs,Strigari2008_dm_dsph,McConnachie2012_dsph}, as predicted by cold-dark-matter-cosmologies~\cite{White1978_galaxies,Springel2005_milsim,Moster2013_galaxies}. This may be seen, for example, by the (dynamical) mass-to-light-ratio ($M_{dyn}/L$) of the dwarf galaxies surrounding the Milky Way~\cite{Mateo1998_dsphs,Strigari2008_dm_dsph}: for intrinsically fainter objects, the expected nearly-constant $M/L$ is (orders of magnitude) larger than for brighter objects (see Fig.~\ref{fig:masstolight}, top panel), which is generally interpreted as an "unseen" mass, or specifically interpreted as dark matter.

Thus, we test for point-like 511~keV emission in 39 dwarf satellite galaxies of the Milky Way, in addition to the diffuse bulge and disk emission, in order to check the dark-matter-511~keV scenario. The point-like assumption is adequate, because typical dark matter density profiles obey an $r^{-\gamma}$ power-law in the inner regions, with $0< \gamma \lesssim 2$~\cite{Burkert1995_dm,Navarro1996_dm,Merritt2006_dm}, and the $\rho^2$-dependence of possible annihilation signals will lead to a very sharply-peaked signal. For example, the angular resolution of SPI of $2.7^{\circ}$ encompasses a physical region of about 1~kpc at a distance of 20~kpc, so that we would see the entire dwarf galaxy at once. In particular, we test for 39 individual emission features with unknown (i.e. to be fitted) fluxes $F_i$, formally adding

\begin{equation}
F_{dwarfs} = \sum_{i=1}^{39} F_i \delta(l-l_i) \delta(b-b_i)\mathrm{,}
\label{eq:fluxes}
\end{equation}

at the respective galaxies' centres $(l_i/b_i)$, to the six-component diffuse base-line model of Siegert et al. (2016a)~\cite{Siegert2016_511,Siegert2016_dsph}. The full list of tested galaxies can be found in Siegert et al. (2016b)~\cite{Siegert2016_dsph}. The dynamical mass to visible light ratio, $\Upsilon_V = M_{dyn}/L_V$, is empirically determined to be $\log(\Upsilon_V) \propto +(0.22\pm0.02)~M_V$, where $M_V$ is absolute visual magnitude, Fig.~\ref{fig:masstolight}. From this estimate, and the definition/conversion of luminosity to V-band magnitude,

\begin{equation}
L_V/L_{\odot} = 10^{-0.4(M_V-M_{V,\odot})}\mathrm{,}
\label{eq:maglumidef}
\end{equation}

we derive a negative proportionality between the dynamical mass and the magnitude, $\log(M_{dyn})$ $\propto$ $-0.2M_V$. The expectation of how the mass-to-511~keV-light-ratio would behave is then determined by 

\begin{equation}
\log(\Upsilon_{511}) := \log\left(\frac{M_{dyn}}{L_{511}}\right) \propto \log\left(\frac{M_{dyn}}{M_{dyn}^2}\right) = -\log(M_{dyn}) = +0.2M_V\mathrm{.}
\label{eq:expetations511}
\end{equation}

It follows that if 511~keV emission is consistently seen in dwarf satellite galaxies, and if it is due to the annihilation of dark matter particles, we would expect the same proportionality for $\Upsilon_{511}$ as for $\Upsilon_V$.

\subsection{511~keV emission from Milky Way satellites}

\begin{figure}[!ht]%
\centering
\includegraphics[width=0.75\columnwidth,trim=2cm 1.5cm 2cm 2cm,clip]{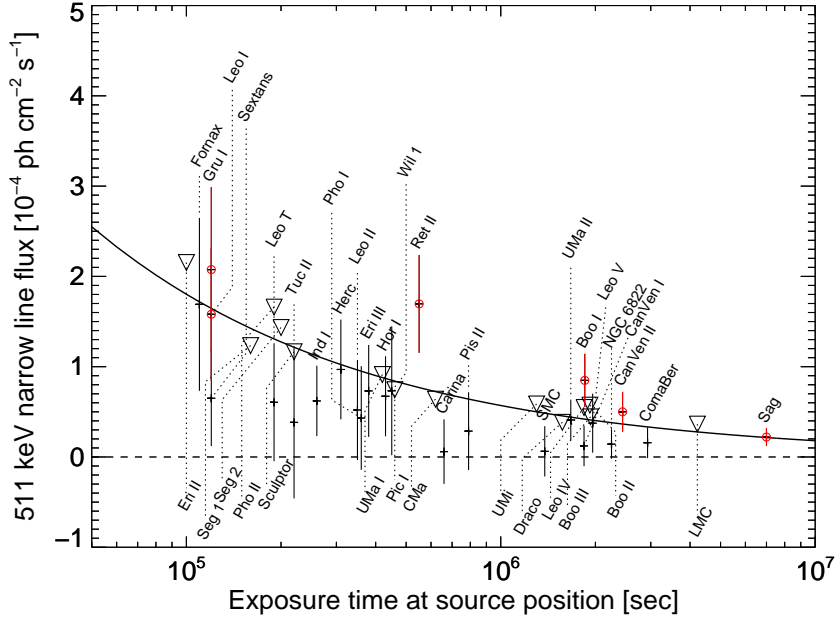}%
\caption{Derived fluxes (crosses) of each satellite galaxy against the exposure time at source position. If a line is not detected or appears negative, a $2\sigma$ upper limit is given (triangle). The solid line represents the $2\sigma$ sensitivity limit for a narrow line (2.15~keV FWHM instrumental resolution) seen with SPI at 511~keV: $5.7 \times 10^{-5} \times \sqrt{10^6 / T_{Exp} [\mathrm{Ms}]}~\mathrm{ph~cm^{-2}~s^{-1}}$. The red circles indicate sources for which the statistical significance is higher than $2\sigma$. See text for details. From Siegert et al. (2016b)~\cite{Siegert2016_dsph}.}%
\label{fig:detectsens}%
\end{figure}

In general, our measurements show more sources with a non-zero 511~keV flux than expected, Fig.~\ref{fig:detectsens}. From pure counting statistics, we would expect 13 sources with more than $1\sigma$ significance above the base-line model (17 seen), 2 sources with more than $2\sigma$ (6 seen), and none with $3\sigma$ or more (1 seen, see Sec.~\ref{sec:retIIsec}). However, it is hard to judge which sources of the sample are real and which are false positive. But we can use the population of non-zero sources to test the expected correlation between $\Upsilon_{511}$ and $M_V$. In Fig.~\ref{fig:masstolight}, $\Upsilon_V$ is shown in the top panel, indicating the "unseen" mass which apparently dominates the faint galaxies. In the bottom panel, $\Upsilon_{511}$ is shown for galaxies which deviate from zero ($1\sigma$-level), and for which dynamical mass estimates are available. The Milky Way is marked by the star symbol in both panels for reference. Apparently, the correlation of $\Upsilon_{511}$ is opposite to $\Upsilon_V$. The measured slope is $\log(\Upsilon_{511}) \propto -(0.25\pm0.11)M_V$, so that this particular dark matter hypothesis can be ruled out on a confidence level of 99.97\% ($3.6\sigma$).  

However, this result should be taken with a grain of salt: Just because we do not consistently see 511~keV emission in those satellite galaxies does not mean that there is no dark matter. But we can be rather confident that the 511~keV emission that we do see is not (dominantly) originating in dark matter, neither in the Milky Way nor in its satellites. But maybe the antimatter that we see is an incarnation of dark matter, rather than its product.

We also provide constraints on the dark matter self-annihilation cross section, $\langle \sigma v \rangle$, based on available J-factors~\cite{Evans2016_Jfactors}, and assuming in-situ $\mathrm{e^+}$-annihilation and negligible formation of positronium atoms. The strongest limit is then derived from the dwarf galaxy Ursa Major II, due to its large J-factor, and is 

\begin{equation}
\langle \sigma v \rangle < 5.6\times10^{-28} \left( \frac{m_{DM}}{\mathrm{MeV}} \right)^2~\mathrm{cm^{3}~s^{-1},}
\label{eq:mdm_limit}
\end{equation}

at a confidence level of $2\sigma$. This constrain is two orders of magnitude above the cross section required to explain the entire Milky Way bulge signal. 

\section{A microquasar in Reticulum II?}\label{sec:retIIsec}

\begin{figure}[!ht]%
\centering
\includegraphics[width=0.75\columnwidth,trim=2cm 1.5cm 2cm 2cm,clip]{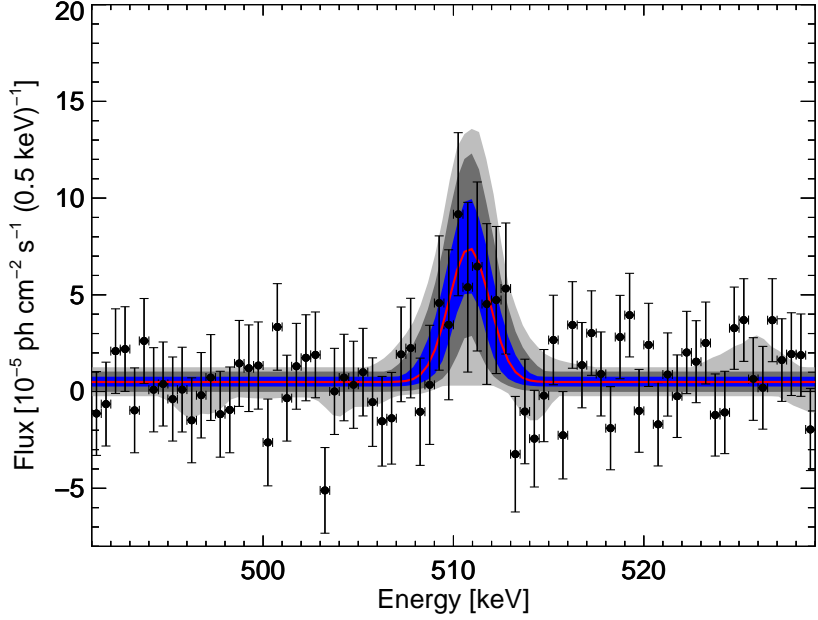}%
\caption{Spectrum from the position of Reticulum II. Shown are the derived data points from the maximum likelihood analysis (black data points) in half-keV binning, together with the best fitting model, a Gaussian on top of a constant offset (red solid line). The $1$, $2$, and $3\sigma$ uncertainty bands of this model are shown in blue, dark grey, and light grey, respectively. See text for details. From Siegert, T. (2017)~\cite{Siegert2017_PhD}.}%
\label{fig:specretII}%
\end{figure}

Among our sample of 39 satellite galaxies, Reticulum II showed up with a $3.1\sigma$ signal. On the one hand, this is a tantalising hint for a bright source of $\mathrm{e^+}$s in this galaxy, on the other hand, the measured 511~keV flux is increasing the evidence against dark matter even further: Reticulum II is an ultra-faint and old dwarf galaxy, located at $(l/b) = (266.3^{\circ}/-49.7^{\circ})$ at distance of 30~kpc, and was discovered by Koposov et al. (2015), using the Dark Energy Survey~\cite{Koposov2015_dsph,Simon2015_dsph_RetII}. In Fig.~\ref{fig:specretII}, the spectrum around 511~keV as measured with SPI is shown. The integrated annihilation line flux is $(1.7\pm0.5)\times10^{-4}~\mathrm{ph~cm^{-2}~s^{-1}}$. The line shows no significant Doppler-shift ($510.8\pm0.4$~keV) nor Doppler-broadening ($1.2\pm0.8$~keV FWHM, astrophysical), though is consistent with the radial velocity of Reticulum II of $\sim70~\mathrm{km~s^{-1}}$. The derived annihilation luminosity in Reticulum II would be about as high as in the entire Milky Way galaxy, $L_{511}(\mathrm{Ret~II}) = (2$-$5)\times10^{43}~\mathrm{e^+~s^{-1}}$. Thus, if this 511~keV signal was purely from dark matter, then the Milky Way bulge should be about 100 times brighter than what is seen with SPI~\cite{Siegert2016_dsph}. This is again strong evidence against a dark matter origin of interstellar $\mathrm{e^+}$s.

Such a large number of $\mathrm{e^+}$s may point to a very special behaviour, source or event in Reticulum II. For example, Ji et al. (2016) measured strong enhancements of neutron-capture elements in stars of the galaxy~\cite{Ji2016_RetII}. They interpret their findings as the result of nucleosynthesis of heavy elements from a single enrichment event only, a neutron star merger. This merger event could also be a source of $\mathrm{e^+}$s if the resulting object was an accreting black hole. In flaring states, such a microquasar could produce large amounts of $\mathrm{e^+e^-}$-pair-plasma~\cite{Siegert2016_V404}, which might be ejected into the interstellar medium of Reticulum II. Such a scenario may indeed be in favour because of complementary X-ray observations with ROSAT. In the vicinity of the baryonic centre of Reticulum II, only 6~arcmin apart, there was (in the 1990s) an unidentified source named 2RXS J033626.8-540215. Between 0.1 and 2.5~keV, the spectrum of this object is strongly absorbed, with a column density ranging between $4$ and $6\times10^{22}~\mathrm{cm^{-2}}$. In order to check for variability of this source and to test the black hole binary hypothesis, we performed a Swift/XRT ToO for 2~ks in July 10, 2016 (Greiner et al. 2016). Between the Swift/XRT range of 0.3 to 10~keV, no significant excess was found near the ROSAT position, so that the source must have significantly (factor of at least 6) decreased in brightness. Consequently, a possible accreting binary black hole might have switched into its quiescent state.

We proposed and were granted additional 1.5~Ms of INTEGRAL observation time in AO15 towards the direction of Reticulum II in order to test for this plausible extragalactic X-ray binary, and also to validate the tentative 511~keV signal in this galaxy.

\section{Conclusions}

We investigated whether the 511~keV $\mathrm{e^+e^-}$-annihilation signal in the Milky Way has its origin in dark matter particle annihilation. Due to morphological similarities, being dominated by a bright and sharply-peaked bulge component, and the smoothness of the emission compared to dark matter halo density profiles, it is expected that neighbouring dwarf satellite galaxies should also show a measurable 511~keV signal. We tested for 39 Milky Way satellites and found a consistent trend which contradicts a dark matter hypothesis, ruling out this particular scenario to more than 99\%. Even though dwarf galaxies might be dominated by a dark matter component, as might be the case for the centre of the Milky Way, the $\mathrm{e^+}$s which are seen to annihilate are predominantly not from dark matter, but rather due to a combination of less-exotic astrophysical sources~\cite{Siegert2016_dsph,Siegert2017_PhD}.

In addition, we found one tentative $3.1\sigma$ annihilation signal from the direction of Reticulum II. This galaxy was also seen in GeV photons~\cite{Geringer-Sameth2015_RetII} in similar way as in the Milky Way centre. Because of indirect evidence for a single enrichment event which could have been a neutron star merger~\cite{Ji2016_RetII}, we connect the strong 511~keV line from Reticulum II to a possible microquasar, ejecting $\mathrm{e^+e^-}$-pair-plasma into the galaxy's interstellar medium. This may also be supported by a source detected with ROSAT near the baryonic centre of Reticulum II in the 1990s, and a non-detection with Swift/XRT in July 2016. Granted future INTEGRAL observations of Reticulum II in AO15 will help to elucidate the scenario of $\mathrm{e^+}$-annihilation in general, its connections to microquasars as significant $\mathrm{e^+}$-producers, and might even reveal a possible link between 511~keV and GeV emission in the Milky Way and its sallites~\cite{Boehm2014_511_GeV}.

\bibliographystyle{bibgen}

\end{document}